\documentstyle[aps,prl,multicol]{revtex}
\begin{document}
\draft
\title{Information-Theoretic Limits of Control}
\author{Hugo Touchette\cite{ht} and Seth Lloyd\cite{sl}}
\address{d'Arbeloff Laboratory for Information Systems and Technology, Department of
Mechanical Engineering, \\ Massachussetts Institute of Technology, Cambridge,
Massachusetts 02139}
\date{\today}
\maketitle

\begin{abstract}
Fundamental limits on the controllability of physical systems are discussed
in the light of information theory. It is shown that the second law of
thermodynamics, when generalized to include information, sets absolute
limits to the minimum amount of dissipation required by open-loop control.
In addition, an information-theoretic analysis of closed-loop control shows
feedback control to be essentially a zero sum game:\ each bit of information
gathered directly from a dynamical systems by a control device can serve to
decrease the entropy of that system by at most one bit additional to the
reduction of entropy attainable without such information (open-loop
control). Consequences for the control of discrete binary systems and
chaotic systems are discussed.
\end{abstract}

\pacs{PACS\ numbers: 05.45.+b, 05.20.-y, 89.70.+c }

\begin{multicols}{2}
Information and uncertainty represent complementary aspects of control.
Open-loop control methods attempt to reduce our uncertainty about system
variables such as position or velocity, thereby increasing our information
about the actual values of those variables. Closed-loop methods obtain
information about system variables, and use that information to decrease our
uncertainty about the values of those variables. Although the literature in
control theory implicitly recognizes the importance of information in the
control process, information is rarely regarded as the central quantity of
interest \cite{bag1}. In this Letter we address explicitely the role of
information and uncertainty in control processes by presenting a novel
formalism for analyzing these quantities using techniques of statistical
mechanics and information theory. Specifically, based on a recent proposal
by Lloyd and Slotine \cite{lloyd1}, we formulate a general model of control
and investigate it using entropy-like quantities. This allows us to make
mathematically precise each part of the intuitive statement that in a
control process, information must constantly be acquired, processed and used
to constrain or maintain the trajectory of a system. Along this line, we
prove several limiting results relating the ability of a control device to
reduce the entropy of an arbitrary system in the cases where (i) such a
controller acts independently of the state of the system (open-loop
control), and (ii) the control action is influenced by some information
gathered from the system (closed-loop control). The results are applied both
to the stochastic example of coupled Markovian processes and to the
deterministic example of chaotic maps. These results not only combine
concepts of dynamical entropy and information in a unified picture, but also
prove to be fundamental in that they represent the ultimate physical
limitations faced by any control systems.

The basic framework of our present study is the following. We assign to the
physical plant ${\cal X}$ we want to control a random variable $X$
representing its state vector (of arbitrary dimension) and whose value $x$
is drawn according to a probability distribution $p(x)$. Physically, this
probabilistic or ensemble picture may account for interactions with an
unknown environment, noisy inputs, or unmodelled dynamics; it can also be
related to a deterministic sensitivity to some parameters which make the
system {\it effectively} stochastic. The recourse to a statistical approach
then allows the treatment of both the unexpectedness of the control
conditions and the dynamical stochastic features as two faces of a single
notion: {\it uncertainty}.

As it is well known, a suitable measure quantifying uncertainty is {\it %
entropy} \cite{shan1,cover1}. For a classical system with a discrete set of
states with probability mass function $p(x)$, it is expressed as 
\begin{equation}
H(X)\equiv -\sum_x\ p(x)\log p(x),  \label{ent1}
\end{equation}
(all logarithms are assumed to the base $2$ and the entropy is measured in
bits). Other similar expressions also exist for continuous state systems
(fine-grained entropy), quantum systems (von Neumann entropy), and
coarse-grained systems obtained by discretization of continuous densities in
the phase space by means of a finite partition. In all cases, entropy offers
a precise measure of disorderliness or missing information by characterizing
the minimum amount of resources (bits) required to encode unambiguously the
ensemble describing the system \cite{shaw1}. As for the time evolution of
these entropies, we know that the fine-grained (or von Neumann) entropy
remains constant under volume-preserving (unitary) evolution, a property
closely related to a corollary of Landauer's principle \cite{land1} which
asserts that {\it only} one-to-one mappings of states, i.e., {\it reversible}
transformation preserving information are exempt of dissipation.
Coarse-grained entropies, on the other hand, usually increase in time even
in the absence of noise. This is due to the finite nature of the partition
used in the coarse-graining which, in effect, blurs the divergence of
sufficiently close trajectories, thereby inducing a ``randomization'' of the
motion. For many systems, the typical average rate of this increase is given
by a dynamical invariant known as the {\it Kolmogorov-Sinai} entropy \cite
{kol1,sin1,alek1}.

In this context, we now address the problem of how a control device can be
used to reduce the entropy of a system or to immunize it from sources of
entropy, in particular those associated with noise, motion instabilities,
incomplete specification of states, and initial conditions. Although the
problem of controlling a system requires more than limiting its entropy, the
ability to limit entropy is a prerequisite to control. Indeed, the fact that
a control process is able to localize a system in definite stable states or
trajectories simply means that the system can be constrained to evolve into
states of low entropy starting from states of high entropy.

To illustrate, in its most simple way, how the entropy of a system can be
affected by external systems, let us consider a basic model consisting of
our system ${\cal X}$ coupled to an environment ${\cal E}$. For simplicity,
and without loss of generality, we assume that the states of ${\cal X}$ form
a discrete set. The initial state is again distributed according to $p(x)$,
and the effect of the environment is taken into account by introducing a
perturbed conditional distribution $p(x^{\prime }|e)$, where $x^{\prime }$
is a value of the state later in time and $e$, a particular {\it realization}
of the stochastic perturbation appearing with probability $p(e)$. For each
value $e$, we assume that ${\cal X}$ undergoes a unique evolution, referred
here to as a {\it subdynamics}, taken to be entropy conserving in analog to
the Hamiltonian time evolution for a continuous physical system: 
\begin{equation}
H(X^{\prime }|e)\equiv -\sum_{x^{\prime }}p(x^{\prime }|e)\log p(x^{\prime
}|e)=H(X).  \label{ent2}
\end{equation}
After the time transition $X\rightarrow X^{\prime }$, the distribution $%
p(x^{\prime })$ is obtained by tracing out the variables of the environment,
and is used to calculate the change of the entropy $H(X^{\prime
})=H(X)+\Delta H$. From the concavity property of entropy, it can be easily
shown that $\Delta H\geq 0$, with equality if and only if (iff) the state $%
{\cal E}$ is perfectly specified, i.e., if a value $e$ appears with
probability one. In practice, however, the environment degrees of freedom
are uncontrollable and the uncertainty associated with the environment
coupling can be suppressed by ``updating'' somehow our knowledge of $X$
after the evolution. One direct way to reveal that state is to imagine a
measurement apparatus ${\cal A}$ coupled to ${\cal X}$ in such a way that
the dynamics of the composed system ${\cal X}+{\cal E}$ is left unaffected.
For this measurement scheme, the outcome of some discrete random variable $A$
of the apparatus is described by a conditional probability matrix $%
p(a|x^{\prime })$ and the marginal $p(a)$ from which we can derive $%
H(X^{\prime }|A)\leq H(X^{\prime })$ with equality iff $A$ is independent of 
$X$ \cite{cover1}. In this last inequality we have used $H(X^{\prime
}|A)\equiv \sum_aH(X^{\prime }|a)p(a)$, and $H(X^{\prime }|a)$ given
similarly as in Eq.(\ref{ent2}).

Now, upon the application of the measurement, one can define the reduction
of entropy of the system conditionally on the outcome of $A$ by $\Delta
H_A\equiv H(X^{\prime }|A)-H(X)$, which, obviously, satisfies $\Delta
H_A\leq \Delta H$, and $H(A)\geq \Delta H-\Delta H_A$. In other words, the
decrease in the entropy of ${\cal X}$ conditioned on the state of ${\cal A}$
is conpensated for by the increase in entropy of ${\cal A}$. This latter
quantity represents information that ${\cal A}$ posses about ${\cal X}$.
Accordingly, the entropy of $X$ given $A$ plus the entropy of $A$ is
nondecreasing, which is an expression of the second law of thermodynamics as
applied to interacting systems \cite{cie1,lloyd2}. In a similar line of
reasoning, Schack and Caves \cite{caves2}, showed that some classical and
quantum systems can be termed ``chaotic'' because of their {\it exponential
sensitivity }to perturbation, by which they mean that the minimal
information $H(A)$ needed to keep $\Delta H_A$ below a tolerable level grows
exponentially in time in comparison to the entropy reduction $\Delta
H-\Delta H_A$.

It must be stressed that the reduction of entropy of ${\cal X}$ discussed so
far is conditional on the outcome of $A$. By assumption, ${\cal X}$ is not
affected by ${\cal A}$; as a result, according to an observer who does not
know this outcome, the entropy of ${\cal X}$ is unchanged. In order to
reduce entropy for all observers unconditioned on the state of any external
systems, a direct dynamical action on ${\cal X}$ must be established
externally by a {\it controller} ${\cal C}$ whose influence on the system is
represented by a set of control actions $x\stackrel{c}{\rightarrow }%
x^{\prime }$ triggered by the controller's state $c$. Mathematically, these
actions can be modelled by a probability transition matrix $p(x^{\prime
}|x,c)$ giving the probability that the system in state $x$ goes to state $%
x^{\prime }$ given that the controller is in state $c$. The specific form of
this {\it actuation} matrix will in general depend on the subdynamics
envisaged in the control process: some of the actions, for example, may
correspond to control strategies forcing several initial conditions to a
common stable state, in which case the corresponding subdynamics is entropy
decreasing. Others can model uncontrolled transitions perturbed by external
or internal noise leading to ``fuzzy'' actuation rules which increase the
entropy of the system. Hence, the systems ${\cal X}$ and ${\cal C}$ need not
in general model a closed system; ${\cal X}$, as we already noted, can also
be affected by external systems (e.g., environment) on which one has usually
no control. However, formally speaking, one can always embed any open-system
evolution in a higher dimensional closed system whose dynamics mimics a
Hamiltonian system. This can be done by supplementing an open system with a
set of {\it ancillary} variables acting as an environment ${\cal E}$ in
order to construct a global volume-preserving transition matrix such that,
when the ancillary variables are traced out, the reduced transition matrix
reproduces the dynamics of the system ${\cal X}+{\cal C}$.

Note that these ancillary variables thus introduced need not have any
physical significance: they are only there for the purpose of simplifying
the analysis of the evolution of the system. In particular, no control can
be achieved through the choice of ${\cal E}$. Within our model, the control
of the system ${\cal X}$ can only be assured by the choice of the control $C$
whereby we can force an ensemble of transitions leading the system to a net
entropy change $\Delta H$. Since the overall dynamics of the system,
controller and environment is Hamiltonian, Landauer's principle immediately
implies that if the controller is initially uncorrelated with the system (%
{\it open-loop} control), a decrease in entropy $\Delta H$ for the system
must be compensated for by an increase in entropy of at least $\Delta H$ for
the controller and the environment \cite{lloyd2}. Furthermore, using again
the concavity property of $H$, it can be shown that the maximum decrease of
entropy achieved by a particular subdynamics of control variable $\hat{c}$
is always {\it optimal} in the sense that no probabilistic {\it mixture} of
the control parameter can improve upon that decrease. Explicitly, we have
the following theorem (we omit the proof which follows simply from the
concavity property.)

{\it Theorem 1.---}For open-loop control, the maximum value of $\Delta H$
can always be attained for a {\it pure} choice of the control variable,
i.e., with $p(\hat{c})=1$ and $p(c)=0$ for all $c\neq \hat{c}$, where $\hat{c%
}$ is the value of the controller leading to $\max \Delta H$. Any mixture of
the control variables either achieves the maximum or yields a smaller value.

From the standpoint of the controller, one major drawback of acting
independently of the state of the system is that often no information other
than that available from the state of ${\cal X}$ itself can provide a
reasonable way to determine which subdynamics are optimal or even accessible
given the initial state. For this reason, open-loop control strategies
implemented independently of the state of the system or solely on its
statistics usually fail to operate efficiently in the presence of noise
because of their inability to react or be adjusted in time. In order to
account for all the possible behaviors of a stochastic dynamical system, we
have to use the information contained in its evolution by considering a {\it %
closed-loop} control scheme in which the state of the controller is allowed
to be correlated to the initial state of ${\cal X}$. This correlation can be
thought as a measurement process described earlier that enables ${\cal C}$
to gather an amount of information given formally in Shannon's information
theory \cite{shan1,cover1} by the {\it mutual information} $%
I(X;C)=H(X)+H(C)-H(X,C),$where $H(X,C)=-\sum_{x,c}p(x,c)\log p(x,c)$ is the 
{\it joint} entropy of ${\cal X}$ and ${\cal C}$. Having defined these
quantities, we are now in position to state our main result which is that
the maximum improvement that closed-loop can give over open-loop control is
limited by the information obtained by the controller. More formally, we have

{\it Theorem 2}.---The amount of entropy $\Delta H_{\text{closed}}$ that can
be extracted from any dynamical system by a closed-loop controller satisfies 
\begin{equation}
\Delta H_{\text{closed}}\leq \Delta H_{\text{open}}+I(X;C),  \label{th2}
\end{equation}
where $\Delta H_{\text{open}}$ is the maximum entropy decrease that can be
obtained by open-loop control and $I(X;C)$ is the mutual information
gathered by the controller upon observation of the system state.

{\it Proof.---}We construct a closed system by supplementing an ancilla $%
{\cal E}$ to our previous system ${\cal X}+{\cal C}$.\ Moreover, let ${\cal C%
}$ and ${\cal E}$ be collectively denoted by ${\cal B}$ with state variable $%
B$. Since the complete system ${\cal X}+{\cal B}$ is closed, its entropy has
to be conserved, and thus $H(X,B)=H(X^{\prime },B^{\prime })$. Defining the
entropy changes of ${\cal X}$ and ${\cal B}$ by $\Delta H=H(X)-H(X^{\prime
}) $ and $\Delta H_B=H(B^{\prime })-H(B)$ respectively, and by using the
definition of the mutual information, this condition of entropy conservation
can also be rewritten in the form $\Delta H=\Delta H_B-I(X^{\prime
};B^{\prime })+I(X;B)$ \cite{lloyd2}. Now, define $\Delta H_{\text{open}}$
as the maximum amount of entropy decrease of ${\cal X}$ obtained in the
open-loop case where $I(X;C)=I(X;B)=0$ (by construction of ${\cal E}$, $%
I(X;E)=0$.) From the conservation condition, we hence obtain $\max \Delta
H=\Delta H_{\text{open}}+I(X;B)$, which is the desired upper bound for a
feedback controller.

To illustrate the above results, suppose that we control a system in a
mixture of the states $\{0,1\}$ using a controller restricted to use the
following two actions 
\begin{equation}
\left\{ 
\begin{array}{l}
c=0:x\rightarrow x^{\prime }=x \\ 
c=1:x\rightarrow x^{\prime }={\sc not}\ x
\end{array}
\right. 
\end{equation}
(in other words, the controller and the system behave like a so-called
`controlled-{\sc not}' gate). Since these actuation rules simply permute the
state of ${\cal X}$, $H(X^{\prime })\geq H(X)$ with equality if we use a
pure control strategy or if $H(X)=H_{\text{max}}=1$ bit, in agreement with
our first theorem. Thus, $\Delta H_{\text{open}}=0$. However, by knowing the
actual value of $x$ ($H(X)$ bit of information) we can choose $C$ to obtain $%
\Delta H=H(X)$, therefore achieving Eq.(\ref{th2}) with equality. Evidently,
as implied by this equation, information is required here as a result of the
non-dissipative nature of the actuations and would not be needed if we were
allowed to use dissipative (volume contracting) subdynamics. Alternatively,
no open-loop controlled situation is possible if we confine the controller
to use entropy-increasing actuations as, for instance, in the control of
nonlinear systems using {\it chaotic} dynamics.

To demonstrate this last statement, let us consider the feedback control
scheme proposed by Ott, Grebogi and Yorke (OGY) \cite{ott1} as applied to
the logistic map 
\begin{equation}
x_{n+1}=rx_n(1-x_n),\quad x\in [0,1],  \label{lgm}
\end{equation}
(the extension to more general systems naturally follows). The OGY method,
specifically, consists of applying to Eq.(\ref{lgm}) small perturbations $%
r\rightarrow r+\delta r_n$ according to $\delta r_n=-\gamma (x_n-x^{*})$,
whenever $x_n$ falls into a region $D$ in the vicinity of the target point $%
x^{*}$. The gain $\gamma >0$ is chosen so as to ensure stability \cite
{schut1}. For the purpose of chaotic control, all the accessible control
actions determined by the values of $\delta r_n$, and thereby by the
coordinates $x_n\in D$, can be constrained to be entropy-increasing for a
proper choice of $D$, meaning that the Lyapunov exponent $\lambda (r)$
associated with any actuation indexed by $r$ is such that $\lambda (r)>0$ 
\cite{note3}. Physically, this implies that {\it almost} any initial uniform
distribution for $X$ covering an interval of size $\varepsilon $ ``expands''
by a factor $2^{\lambda (r)}$ on average after one iteration of the map with
parameter $r$ \cite{vito1}. Now, for an open-loop controller, it can be
readily be shown in that case that no control of the state $x$ is possible;
without knowing the position $x_n$, a controller merely acts as a
perturbation to the system, and the optimal control strategy then consists
of using the smallest Lyapunov exponent available so as to achieve $\Delta
H_{\text{open}}=-\lambda _{\min }<0$. Following theorem 2, it is thus
necessary, in order to achieve a controlled situation $\Delta H>0$, to have $%
I(X;C)\geq \lambda _{\min }$ using a measurement channel characterized by an
information capacity \cite{cover1} of at least $\lambda _{\min }$ bit per
use.

In the controlled regime ($n\rightarrow \infty $), this means specifically
that if we want to localize the trajectory generated by Eq.(\ref{lgm})
uniformly within an interval of size $\varepsilon $ using a set of chaotic
actuations, we need to measure $x$ within an interval no larger than $%
\varepsilon 2^{-\lambda _{\min }}$. To understand this, note that an optimal
measurement of $I(X;C)=\log a$ bits consists, for a uniform distribution $%
p(x)$ of size $\varepsilon $, in partitioning the interval $\varepsilon $
into $a$ subintervals of size $\varepsilon /a$. The controller under the
partition then applies the same actuation $r^{(i)}$ for all the coordinates
of the initial density lying in each of the subintervals $i$, therefore
stretching them by a factor $2^{\lambda (r^{(i)})}$. In the optimal case,
all the subintervals are directed toward $x^{*}$ using $\lambda _{\min }$
and the corresponding entropy change is thus 
\begin{equation}
\Delta H_{\text{closed}}=\log \varepsilon -\log 2^{\lambda _{\min
}}\varepsilon /a=-\lambda _{\min }+\log a,  \label{op1}
\end{equation}
which is consistent with Eq.(\ref{th2}) and yields the aforementioned value
of $a$ for $\Delta H=0$. Clearly, this value constitutes a lower bound for
the OGY scheme since not all the subintervals are controlled with the same
parameter $r$, a fact that we observed in numerical simulations \cite{ht2}.

In summary, we have introduced a formalism for studying control problems in
which control units are analyzed as informational mechanisms. In this
respect, a feedback controller functions analogously to a Maxwell's demon 
\cite{leff1}, getting information about a system and using that information
to decrease the system's entropy. Our main result showed that the amount of
entropy that can be extracted from a dynamical system by a controller is
upper bounded by the sum of the decrease of entropy achievable in open-loop
control and the mutual information between the dynamical system and the
controller instaured during an initial interaction. This upper bound sets a
fundamental limit on the performance of any controllers whose designs are
based on the possibilities to accede low entropy states and was proven
without any reference to a specific control system. Hence, its practical
implications can be investigated for the control of linear, nonlinear and
complex systems (discrete or continuous), as well as for the control of
quantum systems for which our results also apply. For this latter topic, our
probabilistic approach seems particularly suitable for the study of quantum
controllers.

The authors would like to thank J.-J.E. Slotine for helpful discussions.
This work has been partially supported by NSERC (Canada), and by a grant
from the d'Arbeloff Laboratory for Information Systems and Technology, MIT.

\end{multicols}
\end{document}